\documentstyle[12pt]{article}

\catcode`\@=11
\long\def\@makefntext#1{
\protect\noindent \hbox to 3.2pt {\hskip-.9pt
$^{{\ninerm\@thefnmark}}$\hfil}#1\hfill}		%CAN BE USED

 \def\@makefnmark{\hbox to 0pt{$^{\@thefnmark}$\hss}}  %ORIGINAL

\def\ps@myheadings{\let\@mkboth\@gobbletwo
\def\@oddhead{\hbox{}
\rightmark\hfil\ninerm\thepage}
\def\@oddfoot{}\def\@evenhead{\ninerm\thepage\hfil
\leftmark\hbox{}}\def\@evenfoot{}
\def\sectionmark##1{}\def\subsectionmark##1{}}

\newcounter{sectionc}\newcounter{subsectionc}\newcounter{subsubsectionc}
\renewcommand{\section}[1] {\vspace{0.6cm}\addtocounter{sectionc}{1}
\setcounter{subsectionc}{0}\setcounter{subsubsectionc}{0}\noindent
	{\bf\thesectionc. #1}\par\vspace{0.4cm}}
\renewcommand{\subsection}[1] {\vspace{0.6cm}\addtocounter{subsectionc}{1}
	\setcounter{subsubsectionc}{0}\noindent
	{\it\thesectionc.\thesubsectionc. #1}\par\vspace{0.4cm}}
\renewcommand{\subsubsection}[1]
{\vspace{0.6cm}\addtocounter{subsubsectionc}{1}
	\noindent {\rm\thesectionc.\thesubsectionc.\thesubsubsectionc.
	#1}\par\vspace{0.4cm}}

\newcounter{appendixc}
\newcounter{subappendixc}[appendixc]
\newcounter{subsubappendixc}[subappendixc]

\renewcommand{\appendix}[1] {\vspace{0.6cm}
        \refstepcounter{appendixc}
        \setcounter{figure}{0}
        \setcounter{table}{0}
        \setcounter{equation}{0}
        \renewcommand{\thefigure}{\Alph{appendixc}.\arabic{figure}}
        \renewcommand{\thetable}{\Alph{appendixc}.\arabic{table}}
        \renewcommand{\theappendixc}{\Alph{appendixc}}
        \renewcommand{\theequation}{\Alph{appendixc}.\arabic{equation}}
        \noindent{\bf Appendix \theappendixc #1}\par\vspace{0.4cm}}

\def\abstracts#1{{
	\centering{\begin{minipage}{30pc}\tenrm\baselineskip=12pt\noindent
	\centerline{\tenrm ABSTRACT}\vspace{0.3cm}
	\parindent=0pt #1
	\end{minipage}}\par}}

\renewenvironment{thebibliography}[1]
	{\begin{list}{\arabic{enumi}.}
	{\usecounter{enumi}\setlength{\parsep}{0pt}
\setlength{\leftmargin 1.25cm}{\rightmargin 0pt}
	 \setlength{\itemsep}{0pt} \settowidth
	{\labelwidth}{#1.}\sloppy}}{\end{list}}

\newcounter{itemlistc}
\newcounter{romanlistc}
\newcounter{alphlistc}
\newcounter{arabiclistc}

\newcommand{\fcaption}[1]{
        \refstepcounter{figure}
        \setbox\@tempboxa = \hbox{\tenrm Fig.~\thefigure. #1}
        \ifdim \wd\@tempboxa > 6in
           {\begin{center}
        \parbox{6in}{\tenrm\baselineskip=12pt Fig.~\thefigure. #1}
            \end{center}}
        \else
             {\begin{center}
             {\tenrm Fig.~\thefigure. #1}
              \end{center}}
        \fi}

\newcommand{\tcaption}[1]{
        \refstepcounter{table}
        \setbox\@tempboxa = \hbox{\tenrm Table~\thetable. #1}
        \ifdim \wd\@tempboxa > 6in
           {\begin{center}
        \parbox{6in}{\tenrm\baselineskip=12pt Table~\thetable. #1}
            \end{center}}
        \else
             {\begin{center}
             {\tenrm Table~\thetable. #1}
              \end{center}}
        \fi}

\def\@citex[#1]#2{\if@filesw\immediate\write\@auxout
	{\string\citation{#2}}\fi
\def\@citea{}\@cite{\@for\@citeb:=#2\do
	{\@citea\def\@citea{,}\@ifundefined
	{b@\@citeb}{{\bf ?}\@warning
	{Citation `\@citeb' on page \thepage \space undefined}}
	{\csname b@\@citeb\endcsname}}}{#1}}

\newif\if@cghi
\def\cite{\@cghitrue\@ifnextchar [{\@tempswatrue
	\@citex}{\@tempswafalse\@citex[]}}
\def\citelow{\@cghifalse\@ifnextchar [{\@tempswatrue
	\@citex}{\@tempswafalse\@citex[]}}
\def\@cite#1#2{{$\null^{#1}$\if@tempswa\typeout
	{IJCGA warning: optional citation argument
	ignored: `#2'} \fi}}

\def\fnt#1#2{\footnotetext{\kern-.3em
	{$^{\mbox{\sevenrm #1}}$}{#2}}}

 1
 1
 1

\font\tenbf=cmbx10
\font\tenrm=cmr10
\font\tenit=cmti10

\font\ninerm=cmr9

\load{\small}{\it}

\begin{document}

\baselineskip=22pt
\centerline{\tenbf DYNAMICAL CHIRAL-SYMMETRY BREAKING}
\baselineskip=16pt
\centerline{\tenbf IN THE DUAL GINZBURG-LANDAU THEORY}
\vspace{0.8cm}
\centerline{\tenrm SHOICHI SASAKI}
\baselineskip=13pt
\centerline{\tenit RCNP, Osaka University, Ibaraki, Osaka 567, Japan}
\vspace{0.3cm}
\centerline{\tenrm and}
\vspace{0.3cm}
\centerline{\tenrm HIDEO SUGANUMA}
\baselineskip=13pt
\centerline{\tenit RIKEN, Wako, Saitama 351-01, Japan}
\vspace{0.3cm}
\centerline{\tenrm and}
\vspace{0.3cm}
\centerline{\tenrm HIROSHI TOKI}
\baselineskip=13pt
\centerline{\tenit Tokyo Metropolitan University, Hachiohji,
Tokyo 192-03, Japan}
\vspace{0.9cm}
\abstracts{
We study the effect of QCD-monopole condensation on
the dynamical chiral-symmetry breaking by using the dual Ginzburg-Landau
theory of QCD. We formulate the Schwinger-Dyson equation and solve it
numerically.
The large enhancement is found for the chiral-symmetry breaking due to
QCD-monopole condensation, which suggests the close
relation between the color confinement and the chiral-symmetry breaking.
}
\vfil
\rm\baselineskip=14pt
\section{Introduction}
{The lattice QCD simulation showed a remarkable coincidence between
the deconfinement phase transition and the
chiral-symmetry restoration.\cite{Latt93}
It is therefore natural to think the existence of a
strong correlation between confinement and dynamical chiral-symmetry
breaking (D$\chi$SB).
In order to understand these non-perturbative properties, we use the
dual Ginzburg-Landau theory\cite{kanazawa}
as an infrared effective theory of QCD based
on the dual Higgs mechanism.
According to 't Hooft's conjecture,\cite{Hooft}
 QCD is reduced into an abelian gauge theory
with color-magnetic monopoles in the abelian gauge,
which is assumed to be the relevant gauge
for the study of the color confinement in terms of the dual Meissner
effect or monopole condensation. Indeed, the recent
lattice QCD simulations reported the evidence of abelian dominance and
the important role of QCD-monopole condensation for the color
confinement.\cite{{Latt93},{Abel}}

In our previous works, we found that a linear confinement potential is
derived between the static quark-antiquark system as the results of
QCD-monopole condensation in the dual Ginzburg-Landau theory.\cite{SST}
In this paper, we study the role of QCD-monopole condensation on D$\chi$SB
by investigating the Schwinger-Dyson equation in this
theory.\cite{{SST},{SST2}}
}

\section{Dynamical Chiral-Symmetry Breaking}
{The Schwinger-Dyson (SD) equation in the rainbow approximation
 is given by
%
% eq;sde
%
\begin{equation}
S_q^{-1}(p)\;=i{p\kern -2mm /} \;+\;\int {{{d^4k} \over {(2\pi
)^4}}\,}\;\vec Q^2\gamma _\mu \;S_q(k)\;\gamma _\nu \;D^{SC}_{\,\mu \nu
}(p-k)~;~\vec Q^2=C_F{{e^2} \over {N_c+1}}
\label{sde}
\end{equation}
in the Euclidian metric after the Wick rotation in the $k_0$-plane.
Here, $e$ is the gauge coupling constant
and $C_F={{N_c^2-1} \over {2N_c}}$ for $SU(N_c)$.
We take a simple form for the quark propagator, $S_q^{-1}(p)=
i{p\kern -2mm /} - M(p^2)$. By taking the trace of Eq.(\ref{sde}),
an integral equation for the dynamical quark mass $M(p^2)$ is obtained.
Taking the angular average with respect to the direction
of the Dirac string, the trace of the gluon propagator in the Landau gauge is
obtained as
%
% eq;mgp_tr
%
\begin{equation}
D^{SC\;\mu}_{\;\;\mu}(k)={{1} \over k^2} + { 2 \over {k^2+m^2_B}} +
{1 \over \varepsilon} \cdot {4 \over {\varepsilon +
\sqrt {\varepsilon^2 + k^2} }} \left (
{{m^2_B-\varepsilon^2} \over {k^2+m^2_B}} + {\varepsilon^2 \over k^2}
\right ),
\label{mgp_tr}
\end{equation}
which includes the mass of the dual gauge field, $m_B$, and the infrared
cut-off $\varepsilon$.
The mass $m_B$ is generated through the dual Higgs mechanism
when QCD-monopoles are condensed.\cite{{kanazawa},{SST},{SST2}}
The infrared cut-off parameter $\varepsilon$ is introduced for the infrared
screening effect of the strong and long-range correlation due to
the $q$-$\bar q$ pair creation.\cite{{SST},{SST2}}
As for the gauge coupling constant,
we use the Higashijima-Miransky approximation, $\vec Q^2
=4\pi C_F\cdot \alpha_s^{\rm eff}(\max \{ p^2,k^2 \} )$, as the hybrid type of
 the running coupling,\cite{Higashi}
%
% eq;ehig
%
\begin{equation}
\alpha_s^{\rm eff}(p^2)
={{12\pi} \over {(11N_c - 2N_f)\ln \left[ {(p^2+p_c^2)/
{\Lambda _{\rm QCD}^2}} \right]}}
\end{equation}
where $p_c$ approximately divides the momentum scale into the infrared
region and the ultraviolet region, $p_c^2 = {\Lambda^2_{\rm QCD}}
\exp [{{48 \pi^2} \over {e^2}} \cdot {{N_c + 1} \over {11N_c - 2N_f}}]$
.\cite{{SST},{SST2}}
Then, we get the final expression for the SD equation after performing
explicitly the angle integrations,\cite{SST2}
%
% eq;sde2
%
\begin{eqnarray}
M(p^2)
&=& C_F \int_{0}^{\infty} {{{dk^2} \over {4 \pi}}}\;
\alpha_s^{\rm eff}(\max \{ p^2,k^2 \} ) {{M(k^2)}
\over {k^2+M^2(k^2)}} \left( {{k^2} \over {\max \{ k^2,p^2 \} }}
\right. \cr
&& \;\;\;\;\;\;+\;
{{4k^2} \over {k^2+p^2+m^2_B+\sqrt {{(k^2+p^2+m^2_B)}^2-4k^2p^2} }} \cr
&& \;\;\;\;\;\;+\;
{{8k^2} \over {\pi \varepsilon}}\;\int_{0}^{\pi} {d\theta}\;
{{{\sin {\theta}}^2} \over {\varepsilon+\sqrt
{k^2+p^2+\varepsilon^2-2kp \cos {\theta}}}} \cr
&& \;\;\;\;\;\;\;\;\left. \times \left[ {{m^2_B-\varepsilon^2}
\over {k^2+p^2+m^2_B-2kp \cos {\theta}}} +{{\varepsilon^2} \over
{k^2+p^2-2kp \cos {\theta}}} \right] \right),
\label{sde2}
\end{eqnarray}
which is reduced to the usual one of the QCD-like theory in the
ultraviolet limit.\cite{{Higashi},{CSB}}

We solve the SD equation, Eq.(\ref{sde2}), numerically.\cite{SST2}
We show in Fig.1 the dynamical quark mass $M(p^2)$ for several values of
$m_B$, which is proportional to the QCD-monopole condensate.
The quark mass $M(p^2)$ increases with $m_B$, which
means that QCD-monopole condensation provides a large
contribution to D$\chi$SB.\cite{{SST},{SST2}}
(Only a trivial solution exists for small $m_B \leq 0.2{\rm GeV}$).

We also estimate the quark condensate $\langle \bar {q} q\rangle$
and the pion decay constant $f_{\pi}$ under one parameter set
: $e=5.5$, $m_B=0.5{\rm GeV}$ and $\varepsilon =85{\rm MeV}$.
The quark confinement potential (the string tension $k\simeq 1{\rm GeV/fm}$)
and flux tube radius $\simeq 0.4{\rm fm}$ are reproduced with
these parameters.\cite{SST}
Here, the QCD scale parameter is fixed at $\Lambda_{\rm QCD} =200{\rm MeV}$.
We find the dynamical quark mass $M(0) \simeq 354{\rm MeV}$, the quark
condensate $\langle \bar {q} q\rangle =-(222{\rm MeV})^3$
and the pion decay constant
$f_{\pi}=88{\rm MeV}$.\cite{SST2}
Thus, the dual Ginzburg-Landau theory seems to give good results
on both the quark confining potential and D$\chi$SB semi-quantitatively.

In conclusion, we find that D$\chi$SB is
enhanced by QCD-monopole condensation, which leads to the color confinement.
This result indicates
the evidence for the close relation between the color confinement
and D$\chi$SB through QCD-monopole condensation
in the dual Ginzburg-Landau theory.
}
\vspace{0.4cm}
\input epsf
\centerline{\epsfbox{fig1.ps}}
\vspace{0.4cm}
%\fcaption{The dynamical quark mass $\small {\it M}({\it p}^2)$
%as a function of the Euclidian
%momentum squared $\small {\it p}^2$ for $\small \it {m_B}$
%= 0.3, 0.4 and 0.5 GeV. The other
%parameters are $\small \it e$ = 5.5 and $\small \it{\varepsilon}$ = 85MeV,
%while the QCD scale parameter is fixed
%at $\small {\Lambda_{\rm QCD}}$ = 200MeV.
%}
\fcaption{: The dynamical quark mass $M(p^2)$ as a function of the Euclidian
momentum squared $p^2$ for $m_B=0.3, 0.4$ and
$0.5{\rm GeV}$. The other
parameters are $e=5.5$ and $\varepsilon =85{\rm MeV}$, while the QCD scale
parameter is fixed at $\Lambda_{\rm QCD}=200{\rm MeV}$.}

\end{document}